
\input jnl
\rightline {NSF-ITP-09-92}
\rightline {YCTP-P06-92}
\rightline {Imperial/TP/91-92/18}
\rightline {CMU-HEP92-03}
\rightline {February, 1992}
\def\tem{T_{\rm em}}
\def\tm{T_{\rm M}}
\def\timeem{t_{\rm em}}
\def\mm{m_{\rm M}}
\def\mpl{M_{\rm Pl}}
\def\u1{U(1)_{\rm em}}
\title {HOW EFFICIENT IS THE LANGACKER-PI MECHANISM\\
OF MONOPOLE ANNIHILATION?}

\author
{R. Holman${}^{a,b}$, T.W.B. Kibble${}^{a,c}$ and Soo-Jong Rey${}^{a,d,*}$}

\affil {${}^a$Institute for Theoretical Physics, University of California,
Santa Barbara CA 93106\\
${}^b$Physics Department, Carnegie-Mellon University, Pittsburgh PA 15213\\
${}^c$Blackett Laboratory, Imperial College, London SW7 2BZ, UK\\
${}^d$Center for Theoretical Physics, Yale University, New Haven CT 06511}

\abstract
We investigate the dynamics of monopole annihilation by the Langacker-Pi
mechanism. We find that considerations of causality,
flux-tube energetics and the
friction from Aharonov-Bohm scattering suggest that
the monopole annihilation is most efficient
if electromagnetism is spontaneously broken at the lowest
temperature ($\tem \approx 10^6$~GeV) consistent with not having the monopoles
dominate the energy density of the universe.
\vskip1cm
\centerline{\sl submitted to Physical Review Letters}
\vskip1cm
\noindent * Yale-Brookhaven SSC Fellow

\endtopmatter

As is well known, all grand unified theories (GUT's) must of necessity give
rise
to 't Hooft-Polyakov magnetic monopole solitons\refto{monopoles}.
As a practical
matter, these will arise whenever a $U(1)$ subgroup appears after spontaneous
symmetry breaking (a more general criterion involves the second homotopy group
of the vacuum manifold\refto{vilenkinrev}).

{}From a cosmological viewpoint, these monopoles are disastrous. They have a
mass $\mm \sim M_{\rm GUT} \sim 10^{16}\,\rm{GeV}$ and since they are
created via the misalignment of the Higgs fields in different horizon
volumes\refto{kibble}, we expect to have at least one monopole per
horizon at the time
of the GUT phase transition giving rise to the monopoles. These two facts then
lead us to the conclusion that the universe would have become monopole
dominated
long ago and recollapsed shortly thereafter\refto{monopoleproblem}.

Historically, the monopole problem was an important factor in arriving at the
inflationary universe scenario. Indeed, with an appropriate amount of
supercooling (as in the case of a first order phase transition), the monopole
number density could be diluted away. However, there are other solutions to the
monopole problem. In particular, Langacker and Pi\refto{langackerpi} proposed
such a solution some time ago. They argued that if the electromagnetic gauge
group $\u1$ were broken for a period of time and then restored, then
monopole-antimonopole pairs would become bound by flux tubes and then
annihilate
each other. Recently, there has been a revival of interest in this work from a
variety of standpoints\refto{vilenkin, sriva, weinberg, sher, kephart, turok}.

Our aim in this Letter is to elucidate some points concerning the efficiency of
the Langacker-Pi mechanism and in particular, discuss the issue of when
$\u1$ should be broken. The results of our analysis are rather
surprising (at least to us!): the time $\timeem$ at which $\u1$
is broken should be postponed as long as possible, \ie, until just before the
monopoles begin to dominate the energy density of the universe!

This is rather counter-intuitive; the natural expectation, given the energetics
of the monopole-flux tube system is that the temperature $\tem$ corresponding
to
the time $\timeem$ should be as close to the GUT phase transition temperature
$\tm$ as possible. The reason for this is that the tension in the flux tube is
$\sim \tem^2$. Thus the force between monopoles is stronger for larger $\tem$.
However, this cursory analysis neglects some important factors, such as the
role of Aharonov-Bohm scattering by the flux tube, in determining the
annihilation efficiency. It is to these issues we now turn.

\sl 1. Causality Efficiency: \rm
Let us suppose that $\u1$ is broken spontaneously
at a temperature $\tem$ well below the monopole production scale
$\tm$.
The magnetic monopoles were produced with an initial density
$n_{\rm M} (\tm) \approx {\cal O}(1) \xi^{-3} (\tm)$
, where $\xi(T)$ is the correlation length of the Higgs field at
temperature $T$. While the actual value of $\xi(T)$ depends sensitively on the
nature of the GUT phase transition, we can use causality to bound it above by
the horizon size $2 t(\tm)$, where
$ t(T) \approx 0.03 \mpl/ T^2$ during the radiation
dominated era. This yields the following lower bound on the monopole number
density at creation:
$$
n_{\rm M}(\tm) \ge {\cal O} (10^4) {\tm^6 \over  \mpl^3}.
\eqno (1)
$$
If $\u1$ were broken
immediately right after the GUT phase transition,
there would not be enough monopoles available to be connected by the
flux tubes within a Hubble time scale. On the other hand,
at later times when the Universe cools down to a temperature $T$, the \sl
total monopole number inside the horizon \rm grows as
$$
N_{\rm M} (T) \approx {\cal O}(1) \left( \tm \over T \right)^3.
\eqno (2)
$$
The ever increasing total monopole number inside the horizon at temperature
$T <\!\!< \tm$ implies that the flux tube network is easily
formed within a Hubble time scale. For example, when the temperature
$T \approx 10^6$~GeV, at which the Universe starts to become
monopole-dominated, the total monopole number inside a
horizon is $\approx 10^{30}$!

2. \sl Energetic Efficiency: \rm  When
$\u1$ is spontaneously broken, the
flux tube connecting a monopole--anti-monopole pair provides a
linearly increasing confining
potential. The string tension $\mu$ is
$$
\mu \approx \tem^2.
\eqno (3)
$$
If $\tem$ is much less than $\tm$, the motion of the monopole
pair is described by Newton's equation of motion
$$
\mm {d^2 l(t) \over d t^2} = F_{\rm conf} \approx -\tem^2.
\eqno (4)
$$
Here $l(t)$ denotes the monopole--anti-monopole separation (which is the same
as the flux tube length).
The initial separation $l(\timeem)$ should be of the same order of magnitude
as the mean separation distance among the monopoles:
$$\eqalign{
\langle l(\timeem)\rangle  \,\approx & \,\, [n_{\rm M} (\tem)]^{-1/3}
\cr
                             \approx & \,\,  ({\tm \over \tem}) \xi(\tm) \cr
                             \approx & \,\, {\mpl \over 20 \tem \tm}.}
\eqno (5)
$$

The energy stored inside the flux tube is
$$\eqalign {
E_{\rm flux} \equiv & \,\, \mu(\timeem) \langle l(\timeem) \rangle \cr
             \approx & \,\, {\mpl \over  20 \tm} \cdot \tem. }
\eqno (6)
$$
We should mention that if the length in Eq.(5) is long enough so that the
energy
contained in the flux tube is larger than $2\mm$, it becomes energetically
favorable for the tube to break via monopole pair creation. We see from Eq.(5)
that this happens when $\tem > 400 \tm^2 \slash \mpl \approx \tm\slash 25$. In
this case the flux tube may move relativistically and the mean separation after
monopole pair creation by the tube is
$$\eqalign{
\langle l(\timeem)\rangle_{\rm r} \,\approx & \, {20 \tm \over \tem^2} \cr
                              \approx & \,({20 \tm \over \tem}) \cdot
                               ({1 \over \tem}). }
\eqno (7)
$$
We should emphasize that this only happens if $\tem$ is rather close to $\tm$.

{}From Eq.~(4), we find that the characteristic time scale $\tau_{\rm a}$ for
mono
   poles
and antimonopoles to annihilate (assuming an efficient energy
dissipation mechanism; see below) is
$$
\eqalign {
\tau_{\rm a} & \approx \left({\mm \langle l(\timeem) \rangle
\over \tem^2}\right)^{1/2} \cr
     & \approx \left({\mpl \over \tem^3}\right)^{1/2}. }
\eqno (8)
$$
Comparing this with the Hubble time scale $\tau_{\rm H} \approx 2 \timeem$, we
find
$$
{\tau_{\rm a} \over \tau_{\rm H}} \approx 30
\left({\tem \over \mpl}\right)^{1/2}.
\eqno (9)
$$
Hence, the monopole annihilation rate becomes larger as
$\tem$ becomes lower!

Intuitively, this can be understood as follows.
The energetics argument based on the flux tube string tension effect favors
having $\tem$ as close to $\tm$ as possible.
On the other hand, the formation of a
network of monopoles connected by flux tubes favors lower values of $\tem$
as can be seen from Eq.~(2). This is a direct consequence of the
slowing expansion rate of the Universe.
The two effects compete with each other, but the latter dominates
at lower temperatures. Indeed, using Eq.~(2),
one can rewrite Eq.~(9) as
$$
\left(\tau_a \over \tau_H \right)^3 \approx 3 \times 10^4
\left( \tm \over \mpl\right)^{3/2}
\, {1 \over {\sqrt {N (\timeem)}}}.
\eqno (10)
$$
This clearly shows that the monopole annihilation rate depends only
upon the instantaneous total monopole number within the horizon.

\sl 3. Thermal Fluctuations: \rm  So far, we have not taken into account
the effects of
the thermal bath on the monopoles.
These are important since the thermal energy of monopoles provides transverse
velocity to the flux tubes, and thus nonzero angular momentum to the monopole
pair connected by the flux tube.
First of all, monopoles at a temperature $\tem$ are expected
to be in good thermal contact with the background photons and the ambient
plasma.
Indeed, the strength of monopole-photon interaction is of order unity, and
the cross-section for charged plasma-monopole interactions is correspondingly
${\cal O}(\alpha_{\rm{em}}^{-1})$ larger than that among charged particles.

Thus, the initial kinetic and potential energies of the magnetic monopoles at
temperature $\tem <\!\!< {1 \over 25} \tm$ are
$$\eqalign{
K \approx & \,\,\, \tem, \cr
V \approx & \,\, \tem^2 \langle l (\tem) \rangle \cr
  \approx & \,\, 500 \tem .}
\eqno (11)
$$
The typical transverse momentum of the monopoles due to thermal motion is
$P_\perp (\tem) \approx (20 \tm \tem)^{1/2}$.
Thus, the initial angular momentum of the flux tube-monopole
pair reads
$$\eqalign{
L \approx & \,\langle l(\timeem) \rangle  P_\perp(\timeem)\cr
  \approx & \left({\mpl^2 \over 20 \tm \tem}\right)^{1/2}.}
\eqno (12)
$$
In the absence of friction, energy and angular momentum
conservation lead to a final mean separation
$$
\langle \!\langle  l (\tem) \rangle\! \rangle  \,
\approx \,\, {1 \over 20} \left({\mpl \over \tm}\right)^{1/2} {1\over \tem},
\eqno (13)
$$
in which the double bracket denotes an average with thermal fluctuations
taken into account. It is seen that the final mean separation of the
monopole-pair
is larger by a factor of $100$ than the flux tube thickness ${1 \slash
e \tem}$. At the same time, the final transverse momentum of monopoles
at the above separation is of order ${1 \over 10}
( \mpl \tem)^{1/2} <\!\!< \tm$, showing that the monopoles
are always nonrelativistic.
For relativistic monopoles (\ie, if ${1 \over 25}\tm \le \tem \le \tm$),
the transverse momentum $P_\perp \approx E \approx \mpl \tem / \tm$
and $v_\perp \approx 1$. The flux tubes whose original length was given
by Eq.(7) shrink to a mean separation
$$
\langle\!\langle l(\tem) \rangle\!\rangle_{\rm r}\,
\approx \,\,  \left( 10 \tm \over \tem \right)^{1/2} {1 \over \tem}.
\eqno (14)
$$
They are longer than the flux tube
thickness by a factor of $\ge 3$.

In both the relativistic and the nonrelativistic cases, it is seen that
the final monopole-pair
is separated by a centrifugal barrier due to the angular momentum. Thus
the wave-function overlap
and the annihilation cross-section are exponentially suppressed.

This leads us to a crucial point:
in order for the monopole pair to be confined by the
flux tube and annihilate efficiently, the \sl initial angular momentum  \rm
must be dissipated by friction.

\sl 4. Friction from Aharonov-Bohm Scattering: \rm
There are several mechanisms for dissipating the initial angular momentum:
(1) radiation of long-range gluons and/or
weak gauge bosons, (2) interactions between the magnetic monopole and the
ambient plasma,
and (3) the interaction between the flux tube and the plasma through
Aharonov-Bohm scattering.

The interaction between magnetic monopole and the plasma gives rise to
a friction force $F_{\rm M} (T) \approx \rho (T)
\sigma_{\rm CR} v \approx \tem^2 v$ where $\rho$ is the
background plasma energy density, $\sigma_{\rm CR}$
the Callan-Rubakov\refto{callanrubakov, adavis}
cross-section of the mono\-pole and $v$
the monopole terminal velocity. Thus, the monopole dissipation
rate is
$$
\Gamma_{\rm Mon} \approx \left\{
\eqalign{& \left( \tem \over \mpl\right)^{1/2} \tem \hskip1cm
({\rm nonrelativistic}),\cr
             & \left(\tm \over \mpl\right)  \tem \hskip1.5cm
({\rm relativistic}).}
\right.
\eqno (15)
$$

The monopole dissipation rate from radiation of gluons and
weak gauge bosons is found to be\refto{vilenkinrev}
$$
\Gamma_{\rm rad} \approx {1 \over \alpha} \left( \tem \over  \tm\right)^2
{1 \over \langle \!\langle
l \rangle \!\rangle } \approx \left\{
\eqalign{& {10^2 \tem^2  \over \left(\tm^3 \mpl\right)^{1/2}}\ \tem
            \hskip1cm ({\rm nonrelativistic}), \cr \cr
         & \,\, ({ \tem \over  \tm})^{5/2}\ \tem  \hskip1.5cm
          ({\rm relativistic}).}
\right.
\eqno (16)
$$

The Aharonov-Bohm (AB) scattering\refto{abscattering} arises because the
magneti
   c
field is confined inside the flux tube while the color and the weak gauge field
are not. Due to the fractional electric charges
$Q_u = 2e/3$ and $Q_d = -e/3$ carried by the quarks,
the flux tube connecting the monopoles experiences
nontrivial AB scattering with a cross section
$$ {d \sigma_{\rm AB} \over d \theta} =
{\sin^2 \left({Q_{u,d} \over e} \pi\right) \over
2 \pi k \sin^2 {\theta \over 2}}.
\eqno (17)
$$
This result does not contradict the Dirac quantization condition
as the latter applies to the \sl total sum \rm of color, weak isospin and
electromagnetic quantum numbers\refto{ours}.
The AB dissipation rate is
$$
\Gamma_{\rm AB} \approx {\rho \sigma_{\rm AB} \langle \!\langle
l \rangle \!\rangle v \over E}
            \approx \left\{
\eqalign{& \, \left( \tem \over \tm\right)^{1/2}\ \tem
                      \hskip0.5cm ({\rm nonrelativistic}), \cr
         & \,\,\, { \tm^{3/2}\over \mpl}\ \tem^{1/2} \hskip1.5cm
                     ({\rm relativistic}).}
\right.
\eqno (18)
$$
Thus, we find that radiation dissipation is negligible while monopole-plasma
dissipation is suppressed by a geometrical factor $ \tm / \mpl$
or $ \tem /  \tm$ relative to dissipation due to AB scattering.

{}From Eq.~(18), we find that
$$
{\tau_{\rm AB} \over \tau_{\rm a}} \approx
\left\{
\eqalign{
& \left(\tm \over \mpl\right)^{1/2}
\approx  10^{-2} \hskip0.5cm ({\rm nonrelativistic}),\cr
                         & \left( \tem \over \tm\right)^{1/2}
 \hskip2cm ({\rm relativistic}).
}
\right.
\eqno (19)
$$
AB dissipation is most efficient for nonrelativistic
monopoles, \ie, for $ \tem <\!\!<  \tm$.
Similarly, comparing $\tau_{\rm AB}$ with the Hubble expansion time, we find
$$
{\tau_{\rm AB} \over \tau_{\rm H}} \approx \left\{
\eqalign {& 30 { ( \tm  \tem)^{1/2} \over \mpl}
\hskip1.5cm ({\rm nonrelativistic}), \cr
            &  30 \left(  \tem \over  \tm\right)^{3/2} \hskip1.7cm
({\rm relativistic}).}
\right.
\eqno (20)
$$

{}From Eqs. (19) and (20), we thus come to our main conclusion:
the monopole annihilation by the Langacker-Pi
mechanism is most efficient for the lowest possible
$ \tem <\!\!<  \tm$.

Recall that the Hubble time scale increases as $t \propto T^{-2}$,
which is faster
than the monopole annihilation time. This was responsible for the efficiency
of the annihilation at the lower temperature of EM breaking. We have now
found that the friction due to the AB scattering not only dissipates the
angular
momentum efficiently but also helps monopole annihilation at
lower temperature scales!  The time scales involved in the annihilation
dynamics satisfy the following hierarchy:
$$
\tau_{\rm H} \, >\!\!> \, \tau_{\rm a} >\!\!> \, \tau_{\rm AB}
\eqno (19)
$$
for temperatures $\tem <\!\!<  \tm$, thus explaining why the highest efficiency
for monopole annihilation occurs at the lowest possible temperature.
Of course, the scale $\tem$ cannot be too low since the monopoles will
eventually dominate the energy density of the Universe. With the initial
monopole density given by Eq.~(1), we find that the temperature at which
monopoles dominates energy density of the the Universe
(\ie $ \rho_{\rm M} / \rho_{\rm total} \approx
1$) is $t_{\rm c}^{-1} \approx 10^6$~GeV.  Therefore, we can safely set
the lower bound of $ \tem$ as $ \tem \ge 10^6$~GeV.

In this Letter, we have examined the detailed dynamics of the Langacker-Pi
mechanism. Due to the unusual temperature
dependence of the characteristic time scales as summarized in Eq.~(19),
we find the counter-intuitive result
that the most efficient scenario of monopole annihilation occurs
when $\u1$ is broken just before the monopoles dominate the energy
density of the Universe. The fact that the photon is massive and electric
charge
is spontaneously broken leads us to expect that charge
nonconserving processes may provide novel signatures of
the phase, which should be left over until today. In addition, the
Callan-Rubakov effect\refto{callanrubakov} may provide additional
bayron-asymmetry generation at a relatively low energy
scale\refto{sher, kephart}, and we expect
sizable entropy generation from the monopole and anti-monopole annihilation.
We are currently
investigating these issues, and will report as a separate
publication\refto{ours}. After this work was completed we were informed that
E. Gates, L.M. Krauss and J. Terning\refto{krauss} have
recently studied the monopole annihilation efficiency
using W-condensate flux tubes.

We are grateful for the hospitality of the Institute for Theoretical Physics at
Santa Barbara, where this work was initiated. S.J.R. thanks M. Alford
and S. Coleman for useful discussions.  T.W.B.K. thanks A.C. Davis for helpful
comments.
This research was supported in part by the National Science Foundation
under Grant No. PHY89-04035. R.H. was
supported in part by DOE grant DE-AC02-76ER3066, while S.J.R. was supported
in part by funds from the Texas National Research Laboratory
Commission.

\references
\refis {monopoles} G. 't Hooft, Nucl. Phys. \bf B79\rm, 276 (1974); A.M.
Polyakov, JETP Lett. \bf 20\rm, 194 (1974); see also, S. Coleman,
\sl New Phenomena in Subnuclear Physics, \rm ed. A. Zichichi p. 297
(Plenum, New York, 1977).

\refis {vilenkinrev} A. Vilenkin, Phys. Rep {\bf 121}\rm, 265 (1985).

\refis {kibble}T.W.B. Kibble, J. Phys {\bf A9}\rm, 1387 (1976).

\refis {monopoleproblem}  Ya.B. Zeldovich and M.Y. Khlopov,
Phys. Lett. \bf 79B\rm, 239 (1979); J.P. Preskill, Phys. Rev. Lett.
\bf 43\rm, 1365 (1979).

\refis {langackerpi} P. Langacker and S.-Y. Pi, Phys. Rev. Lett. \bf 45\rm, 1
(1980).

\refis {vilenkin} A. Vilenkin, Phys. Lett. \bf 136B\rm, 47 (1984);
A.E. Everett, T. Vachaspati and A. Vilenkin, Phys. Rev. \bf D31\rm, 1925
(1985).

\refis {sriva} A.F. Grillo and Y. Srivastava, Nuovo Cim. Lett. \bf 36\rm, 579
(1983).

\refis {weinberg} E. Weinberg, Phys. Lett. \bf 126B\rm, 441 (1983);
T.W.B. Kibble and E. Weinberg, Phys. Rev. \bf D43\rm, 3188 (1991).

\refis {sher} V.V. Dixit and M. Sher,
Phys. Rev. Lett. \bf 68 \rm 560 (1992).

\refis {kephart} T.H. Farris, T.W. Kephart, T. Weiler and T.-C. Yuan,
Phys. Rev. Lett. \bf 68\rm, 564 (1992).

\refis {turok} A. Pargellis, N. Turok and B. Yurke, \sl Monopole
Anti-Monopole Annihilation in a Liquid Crystal, \rm PUPT-91-1248 preprint
(1991).

\refis {callanrubakov} C.G. Callan, Phys. Rev. \bf D25\rm, 2141 (1982);
V.A. Rubakov, JETP Lett. \bf 33\rm, 644 (1981).

\refis {adavis} R. H. Brandenberger, A. C. Davis, A. M. Matheson, Phys.
Lett. {\bf 218B} 304 (1989).

\refis {abscattering} Y. Aharonov and D. Bohm, Phys. Rev. \bf 119\rm, 485
(1959); R. Rohm, Ph.D. thesis, Princeton University, 1985 (unpublished);
M. Alford and F. Wilczek, Phys. Rev. Lett. \bf 62\rm, 1071 (1989).

\refis {ours} R. Holman, T.W.B. Kibble, S.-J. Rey, A. Singh, F. Freire,
in preparation (1992).

\refis {krauss} E. Gates, L.M. Krauss and J. Terning, to appear (1992).

\endreferences

\endit
\end